\documentstyle[aps,eqsecnum,pre,preprint,multicol]{revtex}

\begin{document}

\title{Liquid-gas phase behaviour of an argon-like fluid modelled
by the hard-core two-Yukawa potential}

\author{D.~Pini}
\address{Istituto Nazionale di Fisica della Materia and Dipartimento di Fisica,
\\ Universit\`a di Milano, Via Celoria 16, 20133 Milano, Italy} 

\author{G.~Stell}
\address{Department of Chemistry, State University of New York at Stony Brook, 
\\ Stony Brook, NY 11794--3400, U.S.A.}

\author{N.B.~Wilding}
\address{Department of Mathematical Sciences, University of Liverpool, 
\\ Peach Street, Liverpool L69 7ZL, U.K.}

\input epsf

\tighten
\maketitle

\begin{abstract} 

We study a model for an argon-like fluid parameterised in terms of
a hard-core repulsion and a two-Yukawa potential. The liquid-gas phase
behaviour of the model is obtained from 
the thermodynamically self-consistent Ornstein-Zernike approximation
(SCOZA) of H{\o}ye and Stell, the solution of which lends 
itself particularly well to a pair potential of this form. 
The predictions for the critical point 
and the coexistence curve are compared to new high resolution simulation 
data and to other liquid-state theories, including the hierarchical reference 
theory (HRT) of Parola and Reatto. Both SCOZA and HRT deliver results 
that are considerably more accurate than standard integral-equation 
approaches. Among the versions of SCOZA considered, 
the one yielding the best agreement with simulation  successfully predicts 
the critical point parameters to within 1\%. 

\end{abstract}




\newpage


\section{Introduction}

Much attention has been paid in recent years to the hard core Yukawa
(HCY) potential as a model for the pair interactions of fluids
\cite{ROSENFELD}. Interest in the potential is motivated on the one
hand by its relevance to solvent averaged interactions in
polyelectrolytic and colloidal particles and, on the other hand, by its
analytical tractability in the context of liquid state theories such as
the Mean Spherical Approximation (MSA) and the Self Consistent
Ornstein-Zernike Approximation (SCOZA). Recent studies of the HCY
fluid can be found in \cite{CACCAMO95,PINI98} and references therein.

For simple fluids interacting via dispersion forces, however, the bare
HCY potential fails to provide a realistic representation of the 
interactions, which are much better modelled by the Lennard-Jones (LJ)
potential. Unfortunately owing to its mathematical structure, the
latter is less amenable to direct study by the MSA and SCOZA than is
the HCY potential. 
Nevertheless, it has long been appreciated that a LJ-like simple-fluid 
potential 
can be replaced by a potential with a hard core plus a linear combination 
of {\em two} Yukawa tails \cite{SUN,JEDRZEJEK,FOILES81,KONIOR,RUDISILL90} 
with no appreciable 
loss of agreement between the experimental and model equations of state. 
Such a potential permits the theoretical
study of simple fluids whilst retaining the convenient mathematical
properties of the HCY potential. 

\begin{figure}[h]
\setlength{\epsfxsize}{8.0cm}
\centerline{\mbox{\epsffile{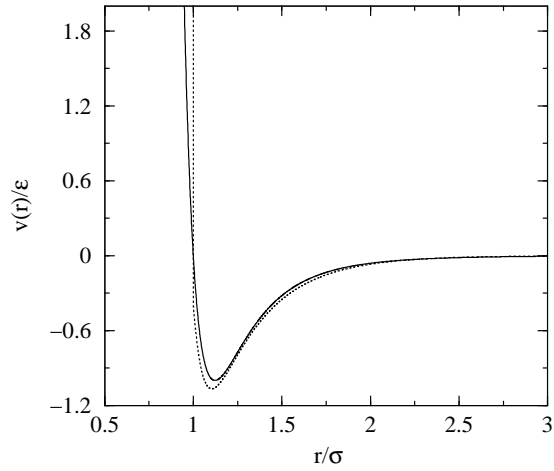}}}
\caption{Pair potentials of the interparticle interaction given by the LJ fluid (dotted curve) and the HC2YF (full curve) 
of Eq.~(\protect\ref{eq:cumpot}) 
with amplitude and range parameters specified in the text.}
\label{fig:potcomp} 
\end{figure}

Several different representations 
of a LJ-like fluid in terms of a hard-core plus two Yukawa fluid (HC2YF)
have been proposed 
in the literature~\cite{SUN,JEDRZEJEK,FOILES81,KONIOR,RUDISILL90}. 
The one we use here is very similar to that recently given by Kalyuzhnyi 
and Cummings \cite{KALY96}:
\begin{equation} 
v(r)=
\left\{
\begin{array}{ll}
\infty \hspace*{7.0cm} & r \le \sigma \, , \\
                                  &                   \\
\displaystyle{\frac{A_{1}\epsilon}{r}}\exp[-z_{1}(r-\sigma)]-
\displaystyle{\frac{A_{2}\epsilon}{r}}\exp[-z_{2}(r-\sigma)] \hspace*{1cm}
& r > \sigma \, .
\end{array}
\right.
\label{eq:cumpot}
\end{equation}
Here $\sigma$ and $\epsilon$ set the length and energy scales of the
model and coincide with the zero and the well depth of the 
LJ potential respectively. The parameters $A_{1}=1.6438\sigma$,
$z_{1}=14.7\sigma^{-1}$, $A_{2}=2.03\sigma$ and $z_{2}=2.69\sigma^{-1}$ 
are chosen
to fit the LJ potential and match the second virial coefficient to that
of the LJ fluid. 
Eq.~(\ref{eq:cumpot}) differs slightly from the original parameterisation by 
Kalyuzhnyi and Cummings insomuch as these authors determined the hard-sphere diameter $\sigma_{\rm HS}$ as a function of temperature via the Barker-Henderson procedure~\cite{BARKER} 
in order to represent the soft-core repulsive interaction
which results from the Weeks-Chandler-Henderson separation 
of the LJ potential~\cite{WCA}. We instead hold the hard-core diameter fixed by setting 
$\sigma_{\rm HS}=\sigma$ for conceptual simplicity and clarity in making comparison with our simulation results.
A comparison of Eq.~(\ref{eq:cumpot}) with the LJ
potential is shown in Fig.~\ref{fig:potcomp}. 
Kalyuzhnyi and Cummings
\cite{KALY96} have used the MSA, Percus-Yevick approximation and
reference hypernetted chain theories to obtain the liquid-gas phase
behaviour of the HC2YF specified by
their parameterisation. Their results showed good agreement
with existing literature data for the LJ fluid. 

In the present work, we extend the work of Kalyuzhnyi and Cummings
\cite{KALY96} by determining the liquid-gas phase behaviour of the
HC2YF using the SCOZA of
H{\o}ye and Stell. Unlike those considered in ref. \cite{KALY96},
this theory is neither mean-field nor mean-spherical like and on the basis 
of a similar study already 
performed for the HCY fluid, it can be expected
to give a superior account of fluid properties in the near-critical
regime. We compare the predictions of the SCOZA for the
the critical point and the coexistence curve 
to those obtained by other approaches, including the renormalization-group 
based hierarchical reference theory (HRT) 
of Parola and Reatto \cite{PAROLA95}, and to new
high resolution simulation results for the HC2YF. We find that one
version of the SCOZA considered provides
a remarkably accurate critical point and coexistence
curve. The critical density and temperature predicted by the theory
agree with the simulation results to within 1\%. 

\section{Theory}

The SCOZA deals with two-body potentials which, like that of 
Eq.~(\ref{eq:cumpot}), consist of a singular hard-sphere repulsion with
diameter $\sigma$ and 
a longer-ranged tail $w(r)$. As is customary in integral-equation theories, 
this approach is based upon the Ornstein-Zernike (OZ) equation linking the
two-body radial distribution function $g(r)$ to the direct correlation 
function $c(r)$. A closed theory is obtained by supplementing the OZ equation 
with an approximate relation involving $g(r)$ and $c(r)$. In its simplest
form, the SCOZA amounts to setting:
\begin{equation}
\left\{
\begin{array}{ll}
g(r)=0                     & r<1 \, , \\
                           & \\
c(r)=K(\rho, \beta)w(r) \mbox{\hspace{0.3 cm}} & r>1  \, , 
\end{array}
\right.
\label{closure}
\end{equation}           
where $\rho$ is the number density of the system, $\beta=1/(k_{\rm B}T)$
is the inverse temperature, and the hard-sphere diameter has been set equal to one. This closure resembles that adopted in the MSA,
except that the amplitude $K$ of the direct correlation function outside
the repulsive core is regarded as an unknown state-dependent quantity, 
to be determined in such a way that consistency between the compressibility 
and the energy route to thermodynamics is enforced. This constraint
amounts to requiring that the reduced compressibility $\chi_{\rm red}$ and
the excess internal energy per unit volume $u$ satisfy the condition:
\begin{equation}
\frac{\partial}{\partial \beta}\left(\frac{1}{\chi_{\rm red}}\right)=
\rho \frac{\partial^2 u}{\partial \rho^2} \, ,
\label{consist}
\end{equation}    
where it is understood that $\chi_{\rm red}$ is obtained by the compressibility 
sum rule as the structure factor $S(k)$ evaluated at $k\!=\!0$, while $u$ is
obtained by the energy equation as the spatial integral of 
the tail interaction $w(r)$ weighted by the radial distribution 
function $g(r)$. 
If the closure~(\ref{closure}) is used for the correlations, the consistency
condition~(\ref{consist}) yields a closed partial differential equation 
(PDE) for the function $K(\rho,\beta)$. In the present case, implementing
this scheme is made simpler by taking advantage of the analytical 
results obtained for the OZ equation with the closure~(\ref{closure})  
when the tail potential $w(r)$ has a two-Yukawa form like that 
of Eq.~(\ref{eq:cumpot}). These enable one to obtain $\chi_{\rm red}$ 
as a function of $\rho$ and $u$. The function $\chi_{\rm red}(\rho,u)$ 
can then be used in Eq.~(\ref{consist}) by taking $u$ instead of 
$K$ as the unknown quantity. The algebraic manipulations are similar
(although not identical) to those we performed in Ref.~\cite{PINI98},
and are based on the results obtained in Ref.~\cite{TWOYUK}. 
We start by introducing the quantity: 
\begin{equation}
f=(1-\xi)\sqrt{\frac{1}{\chi_{\rm red}}} \, ,
\label{f}
\end{equation}
where $\xi=\pi\rho/6$ is the packing fraction. This can be written as:
\begin{equation}
f=-\frac{(z_{1}^{2}-z_{2}^{2})+4\sqrt{q}\, (\gamma_2-\gamma_1)}
{4\left[ (z_1/\!z_2)\, \gamma_{2}-(z_2/\!z_1)\, \gamma_{1}\right]} -
\frac{z_{1}^{2}-z_{2}^{2}}{z_{1}z_{2}} \,
\frac{\gamma_{1}\gamma_{2}(\gamma_{2}-\gamma_{1})}
{\left[ (z_1/\!z_2) \, \gamma_{2}-(z_2/\!z_1) \, \gamma_{1}\right]^2} \, ,
\label{f2}
\end{equation}
where $q=(1+2\xi)^2/(1-\xi)^2$, and $\gamma_{1}$, $\gamma_{2}$ are 
state-dependent functions. By using Eqs.~(\ref{f}) and (\ref{f2}), 
Eq.~(\ref{consist}) becomes: 
\begin{equation}
\frac{2f}{(1-\xi)^{2}}\left[ 
\left(\frac{\partial f}{\partial \gamma_{1}}\right)_{\!\!\rho,\gamma_2} \!\!
\left(\frac{\partial \gamma_{1}}{\partial u}\right)_{\!\!\rho} \! + \!
\left(\frac{\partial f}{\partial \gamma_{2}}\right)_{\!\!\rho,\gamma_1} \!\!
\left(\frac{\partial \gamma_{2}}{\partial u}\right)_{\!\!\rho}
\right] \left(\frac{\partial u}{\partial \beta}\right)_{\!\!\rho}=
\rho\left(\frac{\partial^{2}u}{\partial \rho^{2}}\right)_{\!\!\beta} \, ,
\label{consist2} 
\end{equation}
where the partial derivatives of $f$ with respect to $\gamma_{1}$, 
$\gamma_{2}$ are straightforwardly obtained from Eq.~(\ref{f2}). 
In order to obtain a closed PDE for $u$, one then needs to express 
$\gamma_{1}$, $\gamma_{2}$ as a function of $\rho$ and $u$. To this end, 
we recall the definitions of $\gamma_{1}$, $\gamma_{2}$~\cite{TWOYUK}: 
\begin{equation}
\gamma_{i}=2-\sqrt{q}-\frac{4+2z_{i}-z_{i}^2}{2(2+z_{i})}\, 
\frac{\tau_{i}\, I_{i}-1}{\sigma_{i}\, I_{i}-1} \mbox{\hspace{1 cm}} (i=1,2),   
\label{gamma}
\end{equation}
where the quantities $\sigma_{i}$, $\tau_{i}$ are known and depend only 
on the inverse range $z_{i}$, and $I_{i}$ is given by:
\begin{equation}
I_{i}=4\pi \rho \int_{1}^{+\infty}\! dr\,  r\exp[-z_{i}(r-1)]\, g(r)
\mbox{\hspace{1 cm}} (i=1,2).
\label{int}
\end{equation}
For the two-Yukawa tail potential of Eq.~(\ref{eq:cumpot}), the integrals
$I_{1}$, $I_{2}$ are straightforwardly related to the internal energy per 
unit volume $u$:
\begin{equation}
u=\frac{1}{2}\, \rho \epsilon \, (A_{1}I_{1}-A_{2}I_{2}) \, .
\label{u}
\end{equation}
If Eq.~(\ref{gamma}) is used to express $I_{i}$ as a function of $\gamma_{i}$ 
and the result is substituted into Eq.~(\ref{u}), one obtains an expression
for $u$ as a function of $\rho$, $\gamma_{1}$, $\gamma_{2}$. 
This can be inverted algebrically to obtain $\gamma_{2}$ as a function of
$\rho$, $\gamma_{1}$, and $u$. 
One is then left with the task of expressing $\gamma_{1}$ in terms of $\rho$
and $u$. If we set:
\begin{eqnarray}
x & = & \sqrt{q}-\frac{z_{1}^{2}}{4\gamma_{1}} \, , \label{x} \\
y & = & \sqrt{q}-\frac{z_{2}^{2}}{4\gamma_{2}} \, , \label{y}
\end{eqnarray}    
we find by applying the results of Ref.~\cite{TWOYUK} that the following 
equation hold:
\begin{eqnarray}
& & A_{2}\, z_{2}^{4}\, \sigma_{1}^{2}(2+z_{1})^{2} 
\left[ 4(2-\sqrt{q}-\alpha_{1})(\sqrt{q}-x)-z_{1}^{2}\right]^{2}
\left\{4(z_{2}^{2}-4y^{2})(y-x)^{2}\right. \nonumber \\
& & \mbox{} \left. -(z_{1}^{2}-z_{2}^{2})
\left[z_{1}^{2}-z_{2}^{2}+4(y^{2}-x^{2})\right] \right\}  \nonumber \\
& - & A_{1}\, z_{1}^{4}\, \sigma_{2}^{2}(2+z_{2})^{2} 
\left[ 4(2-\sqrt{q}-\alpha_{2})(\sqrt{q}-y)-z_{2}^{2}\right]^{2}
\left\{4(z_{1}^{2}-4x^{2})(y-x)^{2}\right. \nonumber \\
& & \mbox{} \left. -(z_{1}^{2}-z_{2}^{2}) 
\left[z_{1}^{2}-z_{2}^{2}+4(y^{2}-x^{2})\right] \right\} = 0  \, ,
\label{msa}
\end{eqnarray}   
where $\alpha_{i}$, $i=1,2$, are known functions of the inverse-range
parameters $z_{i}$.   
Note that this equation does not contain the unknown amplitude 
$K(\rho, \beta)$. By writing $x$ and $y$ as functions of $\gamma_{1}$, 
$\gamma_{2}$ via Eqs.~(\ref{x}), (\ref{y}), and subsequentely $\gamma_{2}$
as a function of $\rho$, $\gamma_{1}$, $u$, we finally obtain an equation
for $\gamma_{1}$ as a function of $\rho$ and $u$. If we indicate the l.h.s.
of Eq.~(\ref{msa}) by $F(x,y,\rho)$, the equation for $\gamma_{1}$ has 
the form:
\begin{equation}
F\{x(\gamma_{1}),y[\gamma_{2}(\rho,\gamma_{1},u)],\rho\}=0  \, .
\label{implicit}
\end{equation}
Eq.~(\ref{consist2}) then becomes:
\begin{equation}
B(\rho, u) \left(\frac{\partial u}{\partial \beta}\right)_{\!\!\rho} =
C(\rho, u) \left(\frac{\partial^{2} u}{\partial \rho^{2}}\right)_{\!\!\beta} 
\, ,
\label{consist3}
\end{equation}  
where $B(\rho,u)$ and $C(\rho,u)$ are given by:
\begin{eqnarray}
B(\rho, u) & = & \frac{2f}{(1-\xi)^{2}}\,
\frac{\partial \gamma_{2}}{\partial u}
\left[\frac{\partial f}{\partial \gamma_{2}} \frac{\partial F}{\partial x}
\frac{\partial x}{\partial \gamma_{1}} -
\frac{\partial f}{\partial \gamma_{1}} \frac{\partial F}{\partial y}
\frac{\partial y}{\partial \gamma_{2}}\right] \, ,  \label{b} \\
C(\rho,u) & = & \rho\left[\frac{\partial F}{\partial y}
\frac{\partial y}{\partial \gamma_{2}} 
\frac{\partial \gamma_{2}}{\partial \gamma_{1}} +
\frac{\partial F}{\partial x} 
\frac{\partial x}{\partial \gamma_{1}}\right] \, ,
\label{c}
\end{eqnarray}
where all the partial derivatives are performed at constant density $\rho$. 
 
We note that in Eq.~(\ref{closure}) there is no hard-sphere contribution
to the direct correlation function outside the repulsive core. 
As a consequence, the treatment of thermodynamics and correlations 
of the hard-sphere gas that come out of Eq.~(\ref{closure}) in the 
high-temperature limit coincide with that of the Percus-Yevick (PY) integral
equation. An improved description of the hard-sphere thermodynamics 
at the level of the well known Carnahan-Starling (CS) equation of state 
is desirable in this context, as the slight inaccuracy in the PY treatment 
of the hard-sphere gas affects not only the high-density behavior of the
system in study, but also the location of its critical point and 
of the coexistence curve. A non-vanishing contribution to $c(r)$ for
$r>1$ due to hard-core part of the interaction could be taken
into account by replacing the expression for $c(r)$ of Eq.~(\ref{closure})
with 
\begin{equation}
c(r)=c_{\rm HS}(r)+K(\rho, \beta)w(r) \mbox{\hspace{1 cm}} 
r > 1 \, ,
\label{closure2}
\end{equation}
where the hard-sphere direct correlation function $c_{\rm HS}(r)$ 
can be determined, for instance, by the Waisman 
parameterisation~\cite{WAISMAN}. This is what has been done 
in Ref.~\cite{PINI98} for the HCY fluid. 
However, in the present case of a two-Yukawa tail potential, this would 
require one to deal with a direct correlation function of three-Yukawa
form for $r>1$. Although such an extension of
our treatment appears to be feasible~\cite{KONIOR}, in order to minimize the
complexity of the computation, we have not pursued it
here. Instead, we assumed 
for $c_{\rm HS}(r)$ outside the core a Yukawa form whose range coincides with 
that of the repulsive contribution to the tail potential, and whose 
density-dependent amplitude $H$ is set so as to give CS 
thermodynamics in the high-temperature limit. 
This procedure does not aim at the most accurate description of the hard-core 
contribution to the correlations possible in the context of the SCOZA. 
But it has the advantage of taking into account the hard-sphere thermodynamics 
beyond the PY level without going beyond the two-Yukawa form of $c(r)$. 
An analogous procedure involving only a single Yukawa term was already
implemented for the HCY fluid~\cite{BOULD}, where
it was found to give results very similar to those 
of the more reliable closure~(\ref{closure2}) both for the thermodynamics 
and the phase diagram. According to this prescription, 
for $r>1$, $c(r)$ has the form:
\begin{equation}
c(r)=\displaystyle{\frac{(H+KA_{1})}{r}}\, \exp[-z_1 (r-1)]
-\displaystyle{\frac{KA_{2}}{r}}\, \exp[-z_2 (r-1)] 
\mbox{\hspace{0.8 cm}} r>1 \, , 
\label{hoye}
\end{equation}
while for $r<1$ the core condition $g(r)=0$ holds as before.   
Here $K$ is the unknown, state-dependent amplitude of Eq.~(\ref{closure}),
which vanishes in the high-temperature limit, while $H$ is a known function 
of the density. As the inverse range of the hard-sphere 
contribution to $c(r)$ is locked to $z_{1}$, the treatment of the hard-sphere 
gas that comes out of Eq.~(\ref{hoye}) clearly lacks 
the virial-compressibility consistency of the Waisman parameterisation.    
However, this does not affect the consistency between 
internal energy and compressibility route upon which SCOZA hinges. 
The manipulations that lead to the PDE~(\ref{consist3}) remain unchanged, 
the only difference being that Eq.~(\ref{msa}) will contain an extra term
related to $H$, which will affect the partial derivatives 
$\partial F/\partial x$, $\partial F/\partial y$ that appear 
in Eqs.~(\ref{b}), (\ref{c}).  

The PDE~(\ref{consist3}) has been integrated numerically. The initial
condition at $\beta=0$ and the boundary conditions at the ends of the density
interval are the same as in Ref.~\cite{PINI98} and will not be detailed here. 
Once $u$ has been obtained by solving Eq.~(\ref{consist3}), integration 
of $u$ with respect to $\beta$ yields the Helmholtz free energy and hence
all the other thermodynamic quantities.

\section{Simulation procedure}

The principal aspects of the simulation and finite-size scaling
techniques employed in this work have previously been detailed
elsewhere in the context of a similar study of the Lennard-Jones fluid.
Accordingly we restrict ourself to a brief summary of the methodology
and refer the reader to reference \cite{WILDING95} for a fuller account.

Grand canonical Monte-Carlo (MC) simulations were performed for the
HC2YF model of Eq.~(\ref{eq:cumpot}). The algorithm used had a 
Metropolis form \cite{FRENKEL} and comprised only particle transfer
(insertion and deletion) steps, leaving particle moves to be performed
implicitly as a result of repeated transfers. The potential was cut at
a radius $r_c=3.0\sigma$, and a standard correction term was applied to
the internal energy to compensate for the trunction \cite{FRENKEL}.  To
simplify identification of particle interactions the periodic
simulation space of volume $L^3$ was partitioned into $m^3$ cubic
cells, each of side the cutoff $r_c$. This strategy ensures that
interactions emanating from particles in a given cell extend at most to
particles in the $26$ neighbouring cells. 

System sizes having $m=3,4,5,6,7$ and $8$ were studied, corresponding
(at coexistence) to average particle numbers of approximately
$230,540,1050,1750,2900$ and $4500$ respectively. For the $m=3,4,5$ and
$6$ system sizes, equilibration periods of $10^5$ Monte Carlo transfer
attempts per cell (MCS) were utilised, while for the $m=7$ and $m=8$
system sizes up to $2\times10^6$ MCS were employed. Sampling
frequencies ranged from $20$ MCS for the $m=3$ system to $250$ MCS for
the $m=8$ system. The total length of the production runs was also
dependent upon the system size. For the $m=3$ system size, $1\times
10^7$ MCS were employed, while for the $m=8$ system, runs of up to
$1\times 10^8$ MCS were necessary. 

In the course of the simulations, the observables recorded were the
particle number density $\rho=N/V$ and the energy density $u=E/V$. The
joint distribution $p_L(\rho ,u)$ was accumulated in the form of a
histogram. In accordance with convention \cite{FRENKEL}, we express
all thermodynamic quantities in reduced units: 
$\rho^\ast=\rho\sigma^3$, $u^\ast=u\sigma^3/\epsilon$, 
$T^\ast=k_{\rm B}T/\epsilon$.

Efficient exploration of the phase space was facilitated through use of
the histogram reweighting technique \cite{FERRENBERG}. This method
allows histogram accumulated at one set of model parameters to be
reweighted to provide estimates appropriate to another set of
not-too-distant model parameters. Use of the method permits large areas
of phase space to be mapped using only a few simulations performed at
strategic state points. To facilitate study of the subcritical
coexistence region, the multicanonical preweighting technique
\cite{BERG92} was employed. This method employs a biased sampling
technique to overcome the free energy barrier separating the coexisting
phases and thus allow both to be sampled in a single simulation run. 
When combined with histogram reweighting in the manner described in ref.
\cite{WILDING95}, multicanonical preweighting permits an extremely
efficient accumulation of coexistence curve data. 

\begin{figure}[h]
\setlength{\epsfxsize}{8.0cm}
\centerline{\mbox{\epsffile{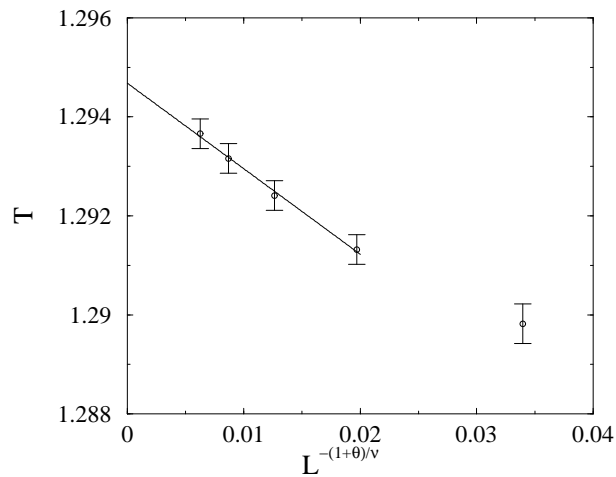}}}
\caption{The apparent reduced critical temperature (as defined by the
matching condition described in the text and in \protect\cite{WILDING95}),
plotted as a function of $L^{-(\theta+1)/\nu}$, with $\theta=0.54$ and
$\nu=0.629$. The extrapolation of the least squares fit to infinite
volume yields the estimate $T_c^\ast=1.295(1)$.}

\label{fig:Tc_inf} 
\end{figure}

The task of estimating the critical point parameters of the model was
performed using finite-size scaling techniques \cite{WILDING95}. In
brief, the strategy is to match the measured ``ordering operator
distribution'' to an independently known universal critical point form
appropriate to the Ising universality class. The ordering operator
itself is defined as ${\cal M}\propto(\rho^\ast+su^\ast)$, where $s$ 
is a non-universal ``field mixing'' parameter, which is finite in the
absence of particle-hole symmetry, and which is chosen to ensure that
$p({\cal M})$ is symmetric in ${\cal M}$\cite{ORKOULAS}. 
For sufficiently large $L$,
the matching should occur at the critical point parameters. For small
$L$, however, the apparent critical temperature obtained by this
matching procedure is subject to systematic errors associated with
corrections to finite-size scaling. To deal with this, we extrapolate
to the thermodynamic limit using the known scaling properties of the
corrections, which are expected to diminish (for sufficiently large
system sizes) like $L^{-\theta/\nu}$ \cite{WILDING95}, where $\theta$
is the correction to scaling exponent and $\nu$ is the correlation
length exponent. The extrapolation has been performed using a least
squares fit to the data for the four largest system sizes. The results
of the extrapolation are shown in Fig.~\ref{fig:Tc_inf}, from which we
estimate $T^\ast_c=1.295(10)$. The associated estimate for the critical
density is $\rho^\ast_c=0.310(1)$ and for the reduced chemical
potential is $\mu^\ast_c=-3.588(30)$.

In addition to the phase coexistence data, we have also measured the
form of the radial distribution function $g(r)$ for a number of
state points, corresponding to reduced temperatures $T^{\ast}=2$, $1.5$, $1$, 
and reduced densities $\rho^{\ast}=0.4$, $0.6$, $0.8$. 
In the following Section, the simulation results for the coexistence curve, 
critical point properties and radial distribution function  
of the HC2YF are compared with the theoretical predictions. 

\section{Results and Discussion}

\begin{figure}[h]
\setlength{\epsfxsize}{8.0cm}
\centerline{\mbox{\epsffile{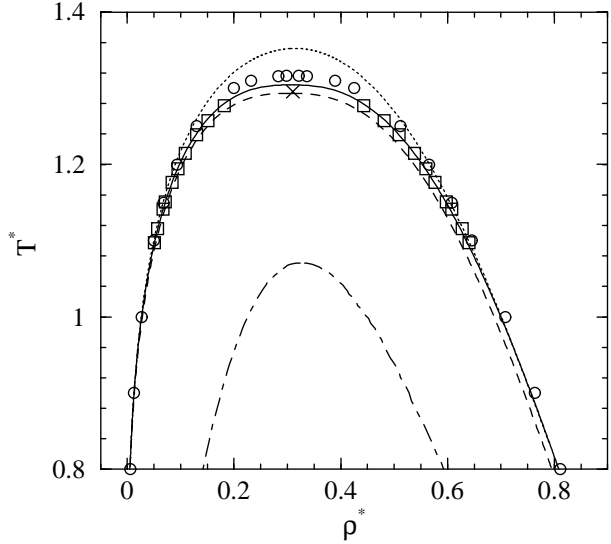}}}
\caption{Coexistence curve of the hard-core two-Yukawa
fluid in the density-temperature plane. Density and temperature are 
in reduced units. Solid line: SCOZA with
closure~(\protect\ref{hoye}) and Carnahan-Starling reference system 
thermodynamics (see text). Dashed line: SCOZA with 
closure~(\protect\ref{closure}) and Percus-Yevick reference system 
thermodynamics. Dotted line: LOGA/ORPA (energy route). Circles: HRT.
Squares: MC simulation performed in this work. Cross: MC critical point.
Dot-dashed line: LOGA/ORPA spinodal curve (compressibility route).}

\label{fig:coex1}
\end{figure}

\begin{figure}
\setlength{\epsfxsize}{8.0cm}
\centerline{\mbox{\epsffile{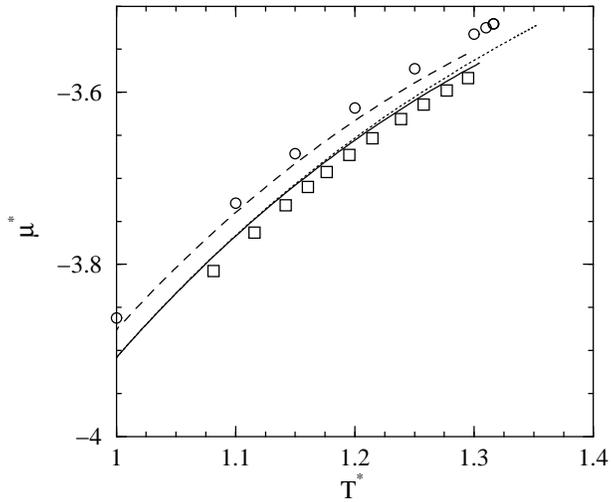}}}
\caption{Coexistence curve in the $T^\ast$-$\mu^\ast$ plane. 
Symbols as in Fig.~\protect\ref{fig:coex1}}
\label{fig:coex2} 
\end{figure}

The SCOZA coexistence curve has been determined by equating the pressure
$P$ and the chemical potential $\mu$ on the low- and high-density branch 
of the subcritical isotherms. As we specified above, these quantities were
obtained via the energy route by integrating $u$ with respect to $\beta$. 
The advantage of doing so is that one does not have to circumvent 
the forbidden region bounded by the spinodal curve, i.e. the locus of 
diverging compressibility, in order to obtain $P$ 
and $\mu$ on the high-density branch. On the other hand, because
of the compressibility--internal energy consistency of the theory, this is
fully equivalent to using a mixed path combining integration of the inverse
compressibility $1/\chi_{\rm red}$ with respect to $\rho$  
and integration of $u$ with respect to $\beta$, as is often
done in calculations based on integral equations~\cite{CACCAMO95}. 
The coexistence curve in the density-temperature and in the 
temperature-chemical potential plane is shown in Figs.~\ref{fig:coex1} 
and~\ref{fig:coex2} respectively.  
Reduced units have been used throughout. 
The SCOZA results obtained both by the
closure~(\ref{closure}) and by the modified version~(\ref{hoye}) are compared 
with the MC data obtained in this work. We have also 
plotted the coexistence curve predicted by HRT and by the energy route
of the lowest-order gamma-ordered approximation (LOGA)~\cite{STELL71},
also known as optimized random-phase approximation (ORPA)~\cite{ANDERSEN72}, 
together with the spinodal curve given by LOGA/ORPA compressibility route. 
It appears that the SCOZA yields
the most accurate determination of the coexistence curve 
among the theories considered here, provided Eq.~(\ref{hoye}) is used
in order to describe hard-sphere thermodynamics at the CS level. 
The predictions for the critical point are compared
in Tab.~\ref{tab:crit}. The error on the SCOZA critical density
and temperature is below $1\%$. Actually, the critical temperature predicted
by SCOZA with closure~(\ref{closure}) and PY hard-sphere thermodynamics is
even closer to the MC result, but this is most likely to be accidental   
since, as stated in Sec.~II, on the basis of our previous calculations on the
HCY potential~\cite{PINI98} we expect that a treatment of the HC2YF 
based on closure~(\ref{closure2})
and a $c(r)$ of three-Yukawa form will in fact deliver results almost identical
to those found here by Eq.~(\ref{hoye}). 
We note that, like the HRT or the energy-route 
LOGA/ORPA, the SCOZA gives a coexistence curve that goes right up to the 
critical point. This feature is not shared by other integral-equation 
approaches, such as the MHNC or the HMSA theories~\cite{CACCAMO95}. 
Both the SCOZA below the critical temperature and the HRT yield non-classical 
critical exponents; for the exponent $\beta_{\rm coex}$ which 
describes the curvature of the coexistence curve the SCOZA gives 
$\beta_{\rm coex}=7/20=0.35$~\cite{HOYE00}, while according the HRT 
$\beta_{\rm coex}\simeq 0.345$~\cite{PAROLA95}, 
the best theoretical estimate being 
$\beta_{\rm coex}\simeq 0.327$~\cite{ZINN83}.  
Moreover, in the SCOZA, as well as in the HRT, the critical point, identified
as the top of the coexistence curve, coincides with that determined 
by locating the divergence of the isothermal compressibility as given by 
the compressibility sum rule. As a consequence, at the critical point one has
both coalescence of the vapor and liquid phases and occurrence of
long-range correlations, as expected. 
Because of the lack of thermodynamic consistency, this is not the case 
with LOGA/ORPA, in which the critical point obtained by the compressibility
route corresponds to the top of the spinodal curve shown 
in Fig.~\ref{fig:coex1}. 

\begin{figure}[h]
\setlength{\epsfxsize}{8.0cm}
\centerline{\mbox{\epsffile{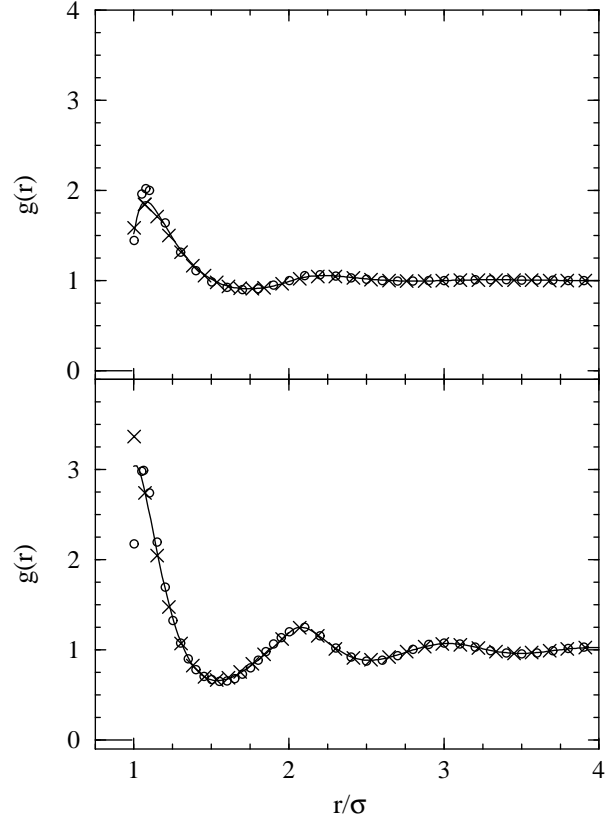}}}
\caption{Radial distribution function of the HC2YF at $T^{\ast}=1.5$, 
$\rho^{\ast}=0.4$ (upper panel) and $T^{\ast}=1$, $\rho^{\ast}=0.8$ 
(lower panel). Solid line: SCOZA with closure~(\protect\ref{closure}). 
Crosses: LOGA/ORPA. Circles: MC simulation. The result of SCOZA 
with closure~(\protect\ref{hoye}) are undistinguishable from those of
LOGA/ORPA.}
\label{fig:corr} 
\end{figure}

We now consider the radial distribution function $g(r)$. 
The SCOZA $g(r)$ has been determined using both closure~(\ref{closure})  
and~(\ref{hoye}). In Fig.~\ref{fig:corr} the SCOZA
and LOGA/ORPA results for two different states are compared 
with MC simulations. We see that, while the overall agreement is satisfactory,
the behavior near contact is not reproduced very well. It is unlikely
that this can be traced back to the slight inaccuracy in the treatment 
of the hard-sphere contribution 
to $c(r)$ for $r>1$ entailed by Eq.~(\ref{closure}) or~(\ref{hoye}). In fact, 
such a contribution is accurately taken into account in LOGA/ORPA, 
which nevertheless does not appear to perform better than SCOZA. 
In particular, using closure~(\ref{hoye}) in SCOZA produces results for $g(r)$ 
which are undistinguishable from those of LOGA/ORPA, at least for the states 
we investigated. Moreover, we checked that for the densities considered here  
the hard-sphere $g(r)$ obtained from closure~(\ref{hoye}) 
in the high-temperature limit is nearly
superimposed to that given by the Waisman parameterisation.
On the other hand, both in SCOZA
and in LOGA/ORPA the contribution to $c(r)$ due to the tail potential
$w(r)$ is bound to follow the profile of $w(r)$ itself for all $r$'s. 
Such a form is best suited for interactions that are slowly varying on
a lengthscale of the order of the particle size, while the two-Yukawa tail
of Eq.~(\ref{eq:cumpot}) changes quite steeply near its minimum. As is well
known, such a problem is usually dealt with by splitting the potential 
according to the WCA prescription~\cite{WCA}, but implementing this procedure
in the SCOZA seems very artificial, as it would hinder the analytical 
tractability of the system, which is the very reason why 
the HC2YF interaction~(\ref{eq:cumpot}) was considered in the first place 
instead of the LJ potential. 

\begin{figure}[h]
\setlength{\epsfxsize}{8.0cm}
\centerline{\mbox{\epsffile{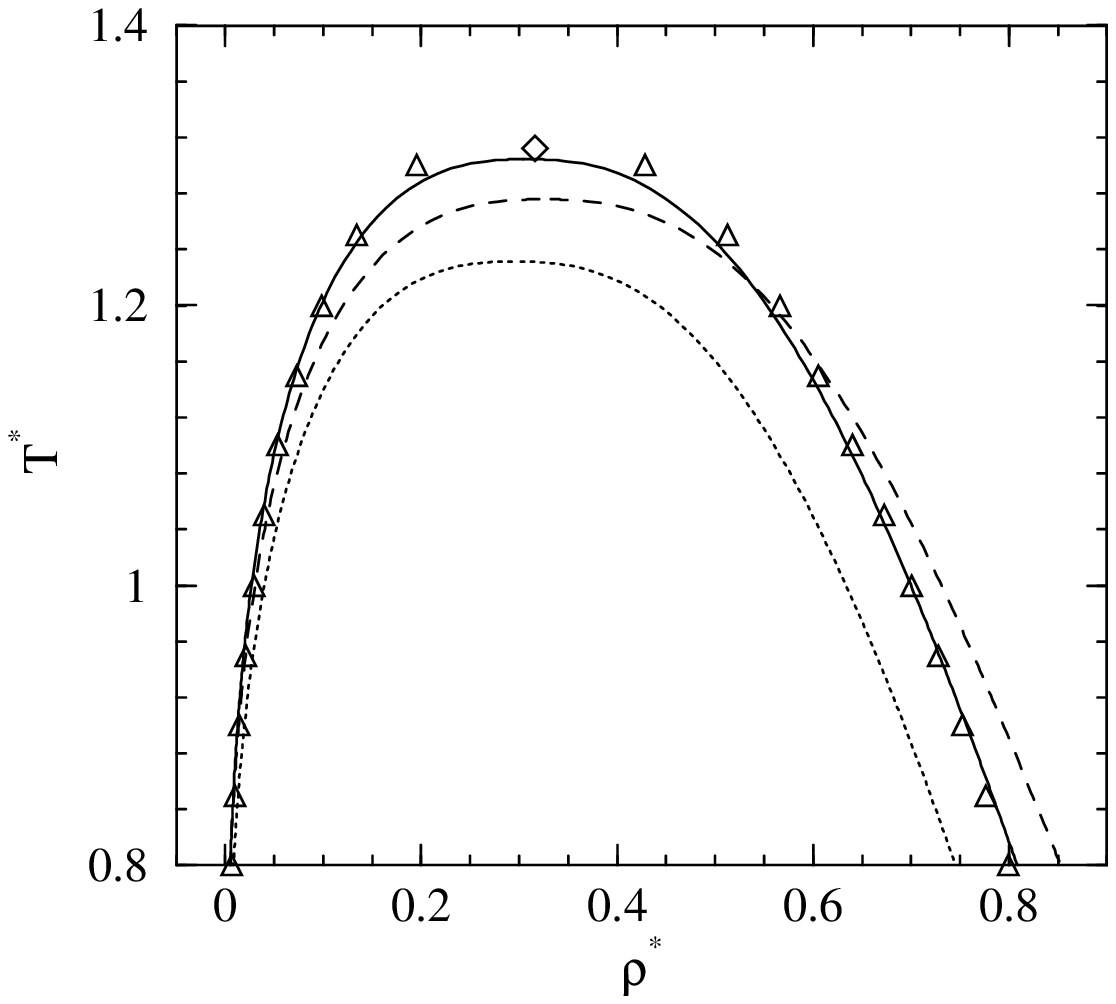}}}
\caption{Comparison between the LJ and the HC2YF coexistence curve  
for different HC2YF parameterisations. Lines: SCOZA results for 
the HC2YF of Eq.~(\protect\ref{eq:cumpot}) with amplitude and range 
parameters by Kalyuzhnyi and Cummings 
(solid line), by Foiles and Ashcroft~\protect\cite{FOILES81} (dotted line), 
and by Sun~\protect\cite{SUN,KONIOR} (dashed line). Triangles: MC simulation 
results for the LJ coexistence curve~\protect\cite{LOTFI}. Diamond: 
MC simulation result for the LJ critical point~\protect\cite{POTOFF98}.} 
\label{fig:comp}
\end{figure}

Since on the basis of the comparison with MC results the SCOZA proves 
to accurately reproduce the coexistence curve of the HC2YF, it could be
worthwhile using it to assess the ability of different HC2YF parameterisations 
of the LJ potential to reproduce the LJ coexistence curve. To this end,
in Fig.~\ref{fig:comp} we have compared the coexistence curve of the 
LJ potential given by MC simulations~\cite{LOTFI} with the SCOZA results
using closure~(\ref{hoye})
for the parameterisations proposed by Sun~\cite{SUN,KONIOR} and by Foiles and
Ashcroft~\cite{FOILES81}, together with that of Eq.~(\ref{eq:cumpot}).
We see that the HC2YF form of Eq.~(\ref{eq:cumpot}), closely resembling 
that of Ref.~\cite{KALY96}, gives the best representation
of the LJ potential among those considered here, at least as far as 
the liquid-vapor phase diagram is concerned. 

In summary, we have presented SCOZA results for the phase diagram and 
the correlations of a two-Yukawa parameterisation of the LJ potential.
These have been compared with the predictions of other theories and with 
new MC simulation data. The comparison shows that the SCOZA provides 
a very accurate coexistence curve, with a critical point that differs 
from the prediction of MC supplemented by finite-size scaling techniques
by less than $1\%$. At the same time, the two-Yukawa potential we studied,
closely resembling that previously proposed by Kalyuzhnyi 
and Cummings~\cite{KALY96}, reproduces satisfactorily 
the LJ coexistence curve. In view of the very good performance of SCOZA
in predicting the liquid-vapor phase diagram of simple fluids, we think that
it could be worthwhile to generalize the solution procedure, so that it will
not be confined anymore to tail potentials of Yukawa form. This could also 
allow one to implement SCOZA with a more sophisticated closure than the
forms ~(\ref{closure}) and (\ref{hoye}) used here.

\acknowledgements

N.B.W. thanks the Royal Society (grant number 19076), the Royal Society
of Edinburgh and the EPSRC (grant no. GR/L91412) for financial support.
D.P. thanks the National Science Foundation for financial support while
visiting Stony Brook. 
G.S. gratefully acknowledges the support of the Division of Chemical Sciences,
Office of Basic Energy Sciences, Office of Energy Research, U.S. Department
of Energy.  

\newpage 
\onecolumn

\begin{table}

\begin{tabular}{ccccccc}  
\makebox[1.2cm]&
\makebox[1.5cm]{\hspace{0.1cm}${\rm MC}^{\dagger}$} & 
\makebox[2cm]{\hspace{0.3cm}${\rm SCOZA}^{\star}$} & 
\makebox[2cm]{\hspace{0.3cm}${\rm SCOZA}^{\bullet}$} &
\makebox[2cm]{\hspace{0.1cm}${\rm HRT}$} &
\makebox[2cm]{\hspace{0.3cm}${\rm LOGA}^{\ddagger}_{\rm en}$} &
\makebox[2cm]{\hspace{0.5cm}${\rm LOGA}^{\diamond}_{\rm comp}$} \\
\hline

\makebox[1.2cm]{$\rho_c^{\ast}$} & 0.310(1) &  0.307 &  0.304 
& 0.310 & 0.314 & 0.328 \\  
\makebox[1.2cm]{$T_c^{\ast}$} & 1.295(1) &  1.304 & 1.293   
& 1.316 & 1.352 & 1.071 \\ 
\end{tabular}
\caption{
Critical density and temperature (in reduced units) for the HC2YF.
$\dagger$:~MC simulation performed in this work.
$\star$:~SCOZA result using 2-Yukawa $c(r)$ of Eq.~(\protect\ref{hoye}).
$\bullet$:~SCOZA result using 2-Yukawa $c(r)$ of Eq.~(\protect\ref{closure}). 
$\ddagger$:~LOGA/ORPA result using energy route. 
$\diamond$:~LOGA/ORPA result using compressibility route.} 

\label{tab:crit} 
\end{table}

\newpage


\end{document}